\documentclass[twocolumn,twoside,slac_two]{revtex4}
\usepackage{graphicx}
\usepackage{fancyhdr}
\def\apj{ApJ}  
\def\aj{AJ}  
\def\apjl{ApJL}  
\def\aap{A\&A} 

\def\mnras{MNRAS} 
\def\pasa{PASA} 

\begin{document}

\title{Unification and physical interpretation of the radio spectra variability patterns in Fermi blazars and jet emission from NLSy1s}

%

\author{E. Angelakis,  L. Fuhrmann, I. Nestoras, C. M. Fromm, R. Schmidt, J. A. Zensus, N. Marchili, T. P. Krichbaum}
\affiliation{Max-Planck-Institut f\"ur Radioastronomie, Auf dem H\"ugel 69, Bonn 53121, Germany}
\author{M. Perucho}
\affiliation{Department d'Astronomia i Astrof\'{i}sica, Universitat de Val\`{e}ncia, C/Dr. Moliner 50, 46100 Burjassot, Val\`{e}ncia, Spain}
\author{H. Ungerechts, A. Sievers, D. Riquelme}
\affiliation{Instituto de Radio Astronomía Milim\'{e}trica, Avenida Divina Pastora 7, Local 20, E 18012, Granada, Spain}
\author{L. Foschini}
\affiliation{INAF - Osservatorio Astronomico di Brera, Italy}

\begin{abstract}
  The {\em F-GAMMA} program is among the most comprehensive programs that aim at
  understanding the physics in active galactic nuclei through the multi-frequency
  monitoring of {\em Fermi} blazars. Here we discuss monthly sampled broad-band radio
  spectra (2.6 - 142\,GHz). Two different studies are presented. (a) We discuss that the
  variability patterns traced can be classified into two classes: (1) to those showing
  intense spectral-evolution and (2) those showing a self-similar quasi-achromatic
  behaviour. We show that a simple two-component model can very well reproduce the
  observed phenomenologies. (b) We present the cm-to-mm behaviour of three $\gamma$-ray
  bright Narrow Line Seyfert 1 galaxies over time spans varying between $\sim1.5$ and 3 years
  and compare their variability characteristics with typical blazars.
\end{abstract}

\maketitle

\thispagestyle{fancy}


\section{BLAZAR VARIABILITY}
Among the most prominent characteristics of blazars, is their intense variability at all
energy bands from radio to $\gamma$-rays and TeV energies. The mechanism producing
variability itself depends on the energy band it refers to and has been long debated. The
``Shock-in-Jet'' model suggested by \cite{Marscher1985ApJ}, is broadly accepted to explain
the variability in terms of shocks propagating in the jet. The basic assumption is that
changes in the particle injection rate, the magnetic field and the bulk Lorentz factor taking place
at the jet base, cause the formation of shocks which consequently go through first {\em
  Compton}, then {\em synchrotron} and finally {\em adiabatic} energy loss phases.  An
alternative model, the ``Internal Shock Model'' proposed by \cite{Spada2001MNRAS},
suggests that energy is channeled into the jet in an intermittent way. ``Plasma shells'',
may have different bulk Lorentz factors and masses, so that faster shells can catch up
with slower ones, collide with them and create relativistic shocks. The shock accelerates
electrons to relativistic energies so that they can emit synchrotron and inverse Compton
radiation. Other models explain the variability geometrically. \citet{1999AnA...347...30V}
for instance, suggested that orbital motion and jet precession, possibly caused by a
binary black hole system, can produce helical jet morphologies which may then cause
brightness and spectral variability through changes in the Doppler factor.

Studies of the variability of the {\em Spectral Energy Distribution} (SED), ideally based
on simultaneous datasets, probe the physics driving the energy production and dissipation
in these systems (e.g. \cite{boettcher2010,boettcher2010HEAD}) and can readily reveal some
physical parameters such as the brightness temperatures, Doppler factors and the jet viewing
angles.

The current report deals with the variability characteristics of AGNs as they are observed
in radio bands at cm to mm wavelengths and has a twofold character. First, we present the
phenomenological classification of the variability patterns followed by selected {\sl
  Fermi} blazars and propose a toy-model which can physically explain this
classification. Secondly, we present the radio behaviour of three {\em Narrow Line Seyfert
  1} galaxies (hereafter NLSy1) which have been detected in $\gamma$-rays by {\sl
  Fermi}/LAT .

\section{THE F-GAMMA PROGRAM}
The potential of the LAT, on-board the {\em Fermi} satellite \citep{2009ApJ...697.1071A},
in the study of AGNs, could be attributed mostly to (a) the covered energy range (from
20\,MeV to $>$300\,GeV) and (b) the pace at which the $4\pi$ $\gamma$-ray sky is being
scanned. Given the broadness of the blazar {\em spectral energy distribution} (SED), the
full potential of such densely sampled $\gamma$-ray light curves can be explored only when
combined with multi-frequency data. Radio monitoring programs are essential in this task
since they trace the behaviour of the relativistic jets where both primary and secondary
emission processes take place (i.e. synchrotron and inverse Compton processes).
\begin{figure*}[]
\begin{minipage}{12pc}
\includegraphics[width=0.67\textwidth,angle=-90]{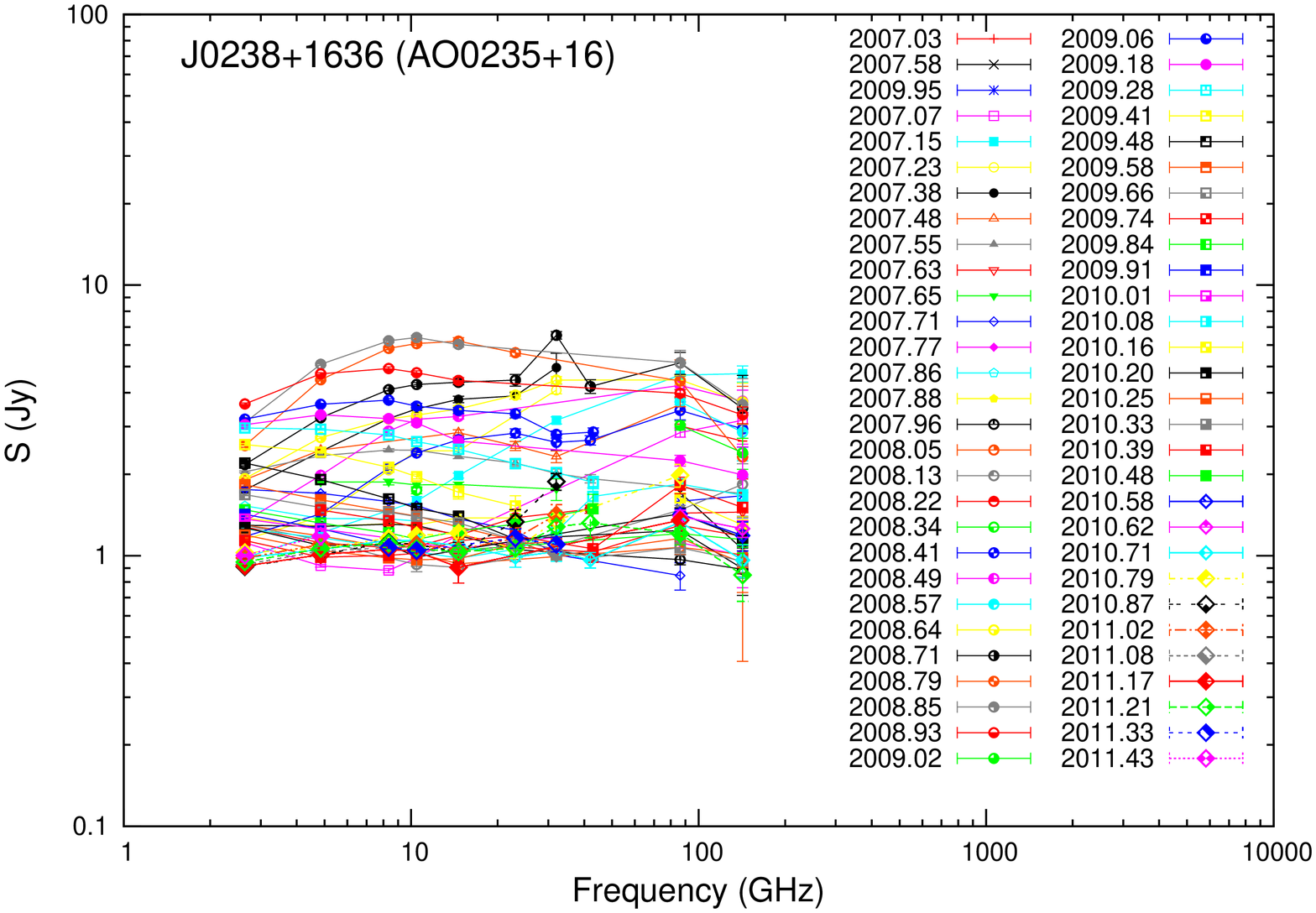}
\caption{\label{fig:t1}Prototype source for variability type 1.}
\end{minipage}\hspace{1pc}\vspace{1pc}%
\begin{minipage}{12pc}
\includegraphics[width=0.67\textwidth,angle=-90]{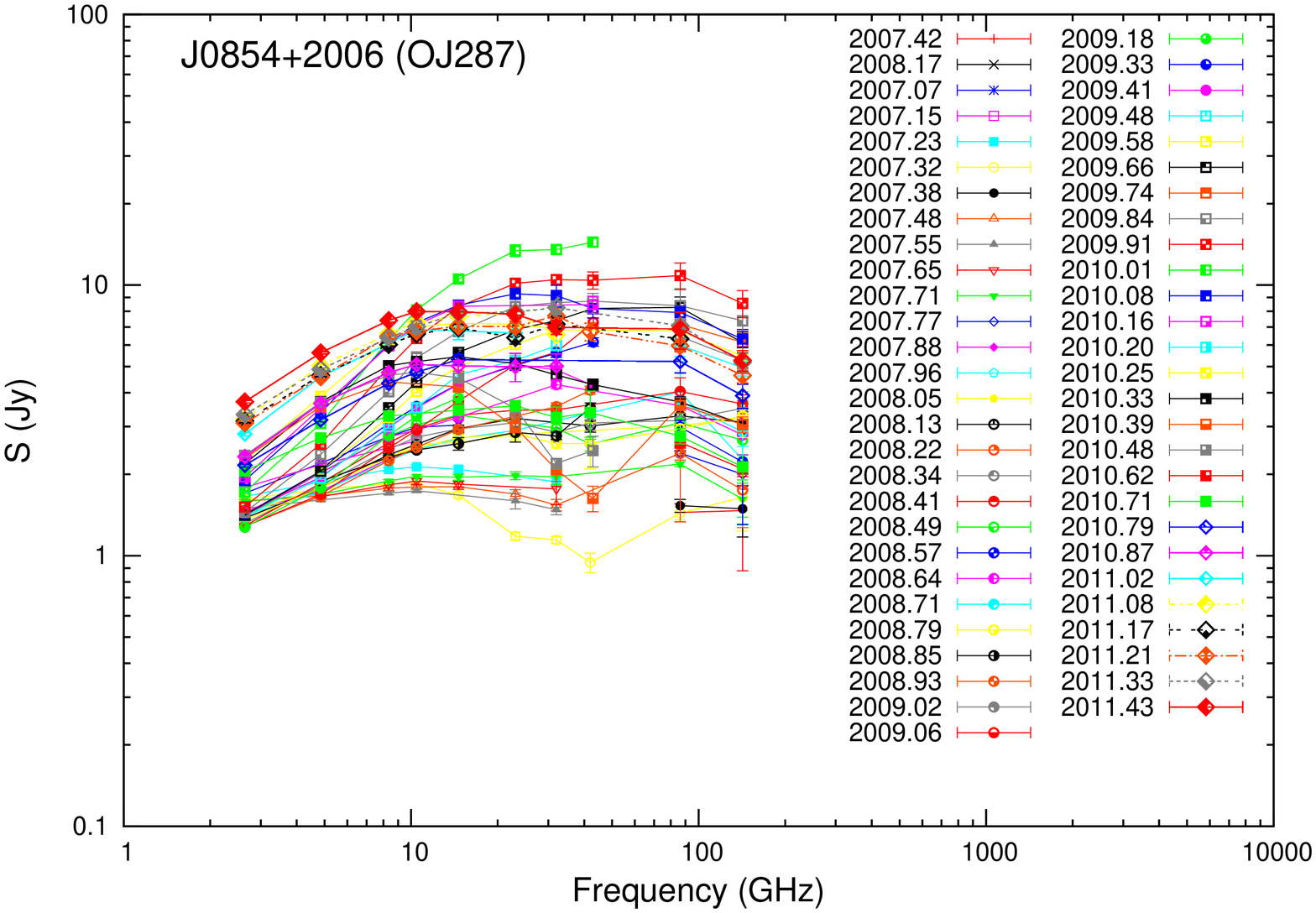}
\caption{\label{fig:t1b}Prototype source for variability type 1b.}
\end{minipage}\hspace{1pc}%
\begin{minipage}{12pc}
\includegraphics[width=0.67\textwidth,angle=-90]{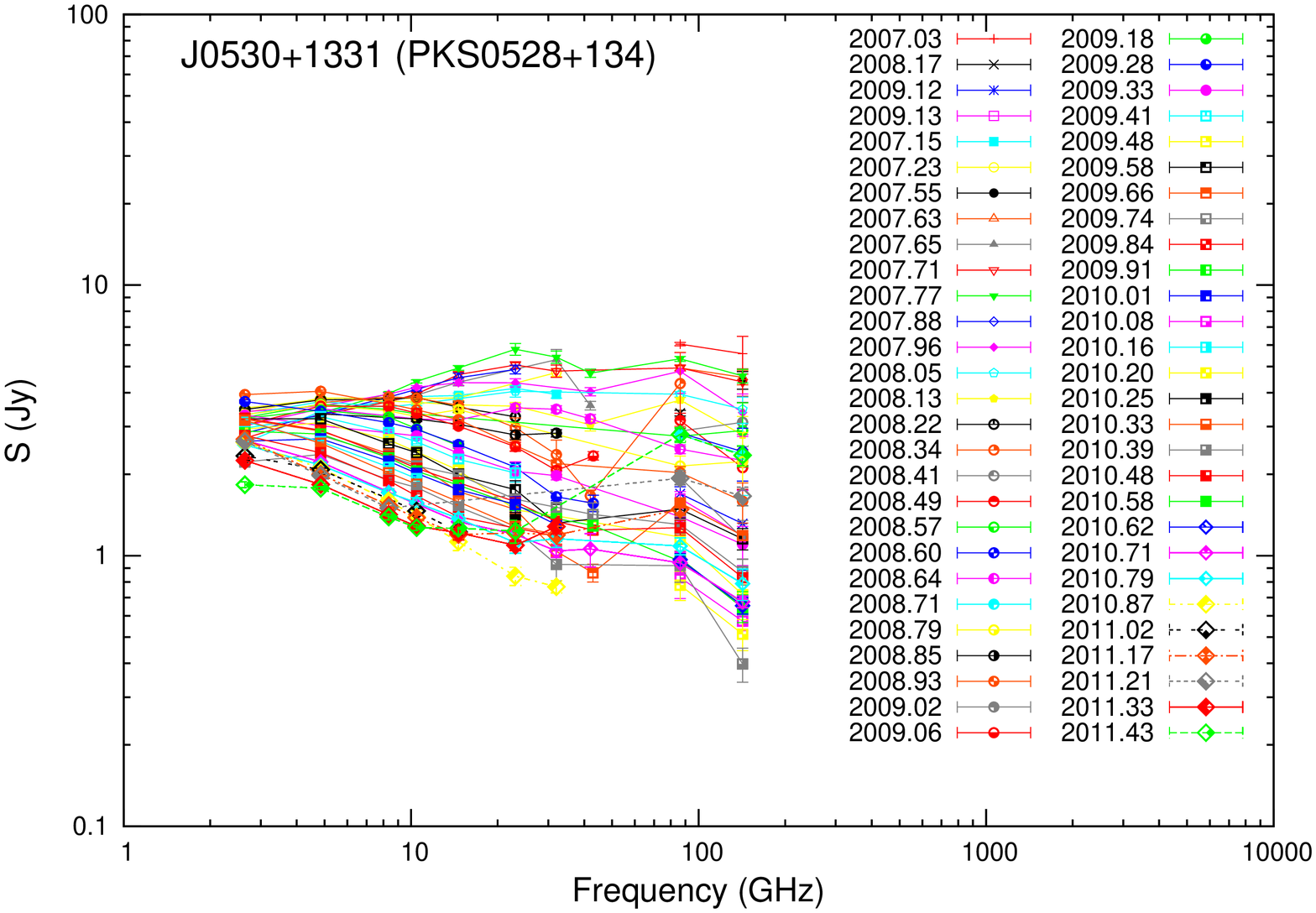}
\caption{\label{fig:t2}Prototype source for variability type 2.}
\end{minipage}\hspace{1pc}%
\begin{minipage}{12pc}
\includegraphics[width=0.67\textwidth,angle=-90]{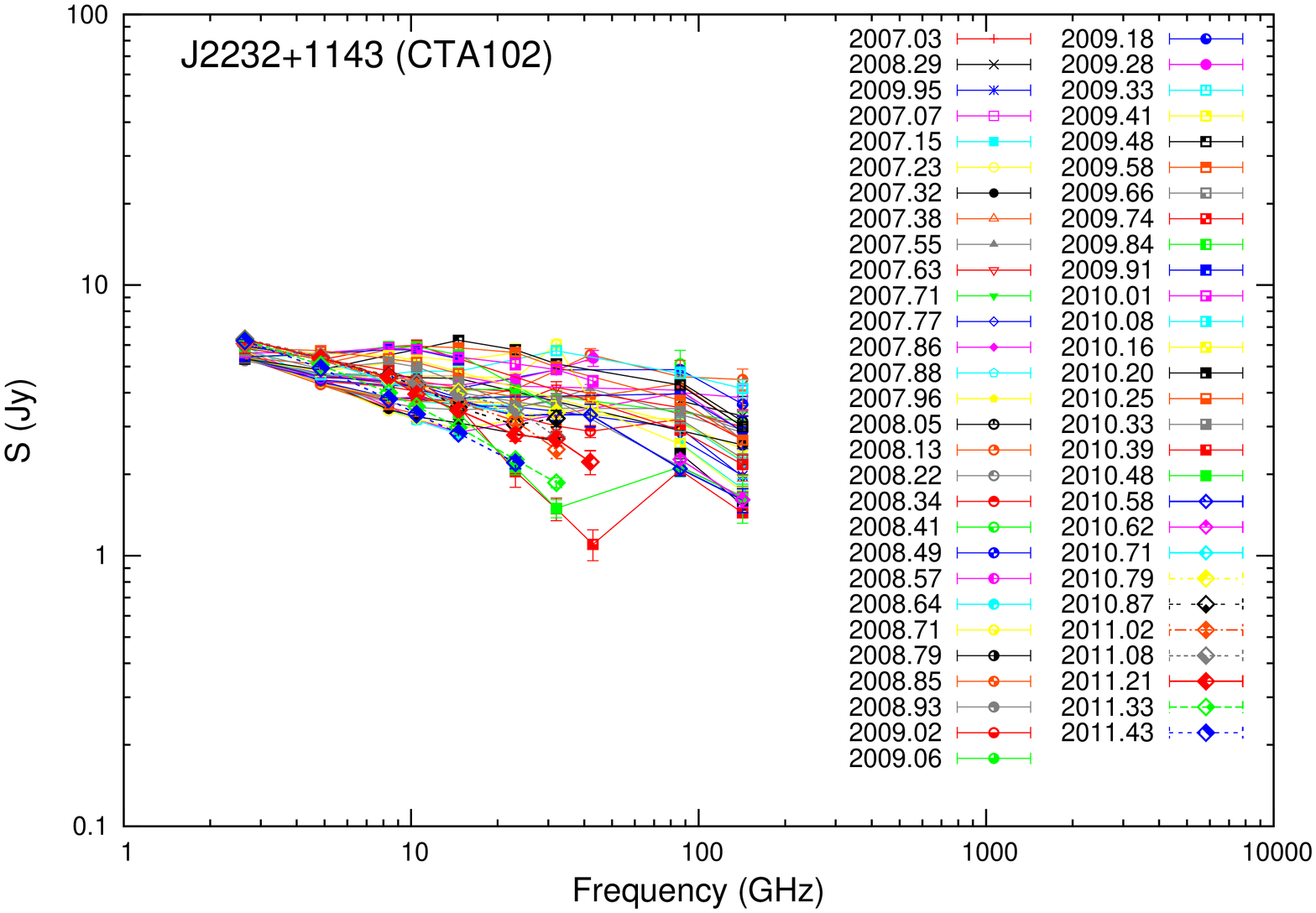}
\caption{\label{fig:t3}Prototype source for variability type 3.}
\end{minipage}\hspace{1pc}\vspace{1pc}%
\begin{minipage}{12pc}
\includegraphics[width=0.67\textwidth,angle=-90]{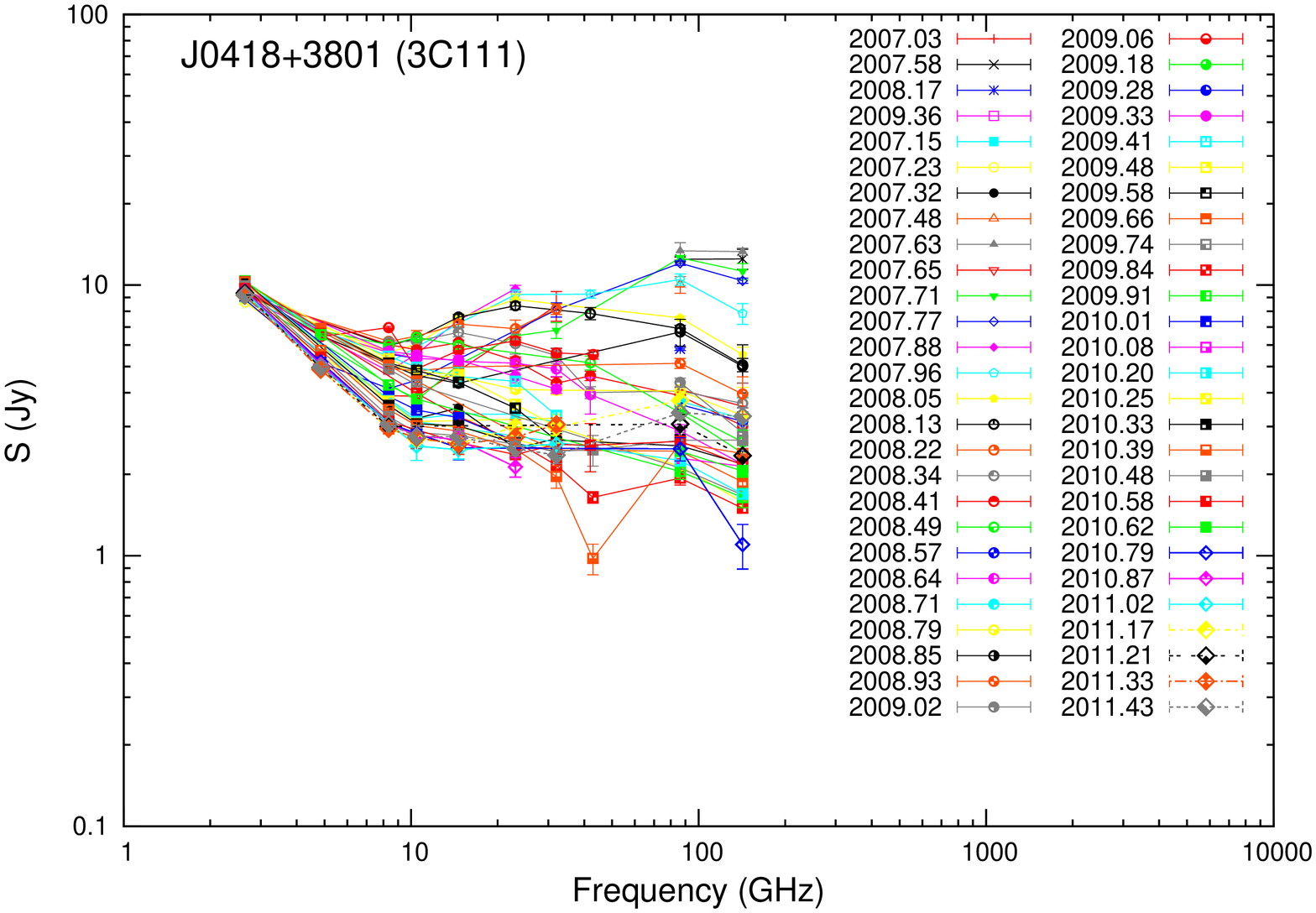}
\caption{\label{fig:t3b}Prototype source for variability type 3b.}
\end{minipage}\hspace{1pc}%
\begin{minipage}{12pc}
\includegraphics[width=0.67\textwidth,angle=-90]{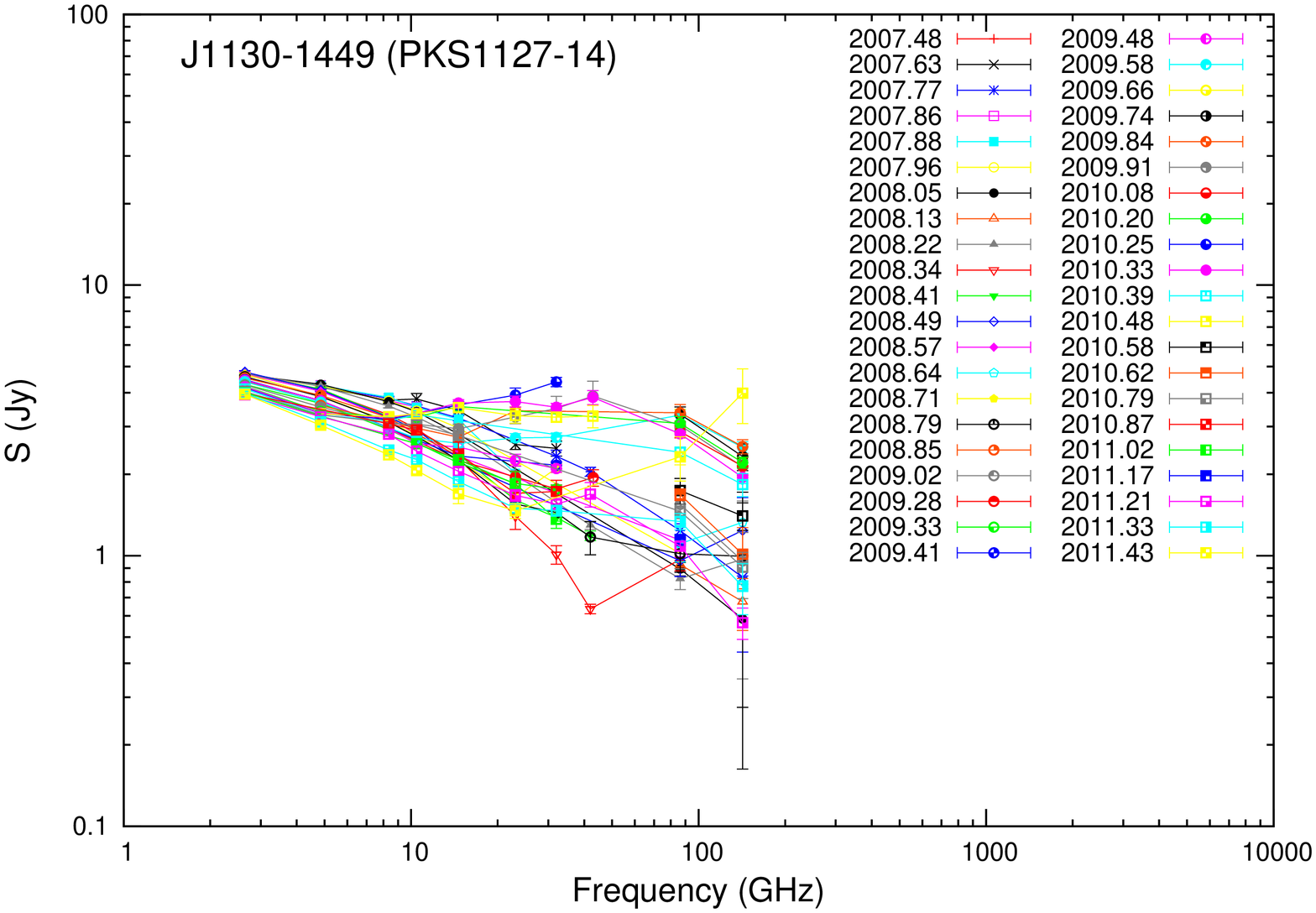}
\caption{\label{fig:t4}Prototype source for variability type 4.}
\end{minipage}\hspace{1pc}%
\begin{minipage}{12pc}
\includegraphics[width=0.67\textwidth,angle=-90]{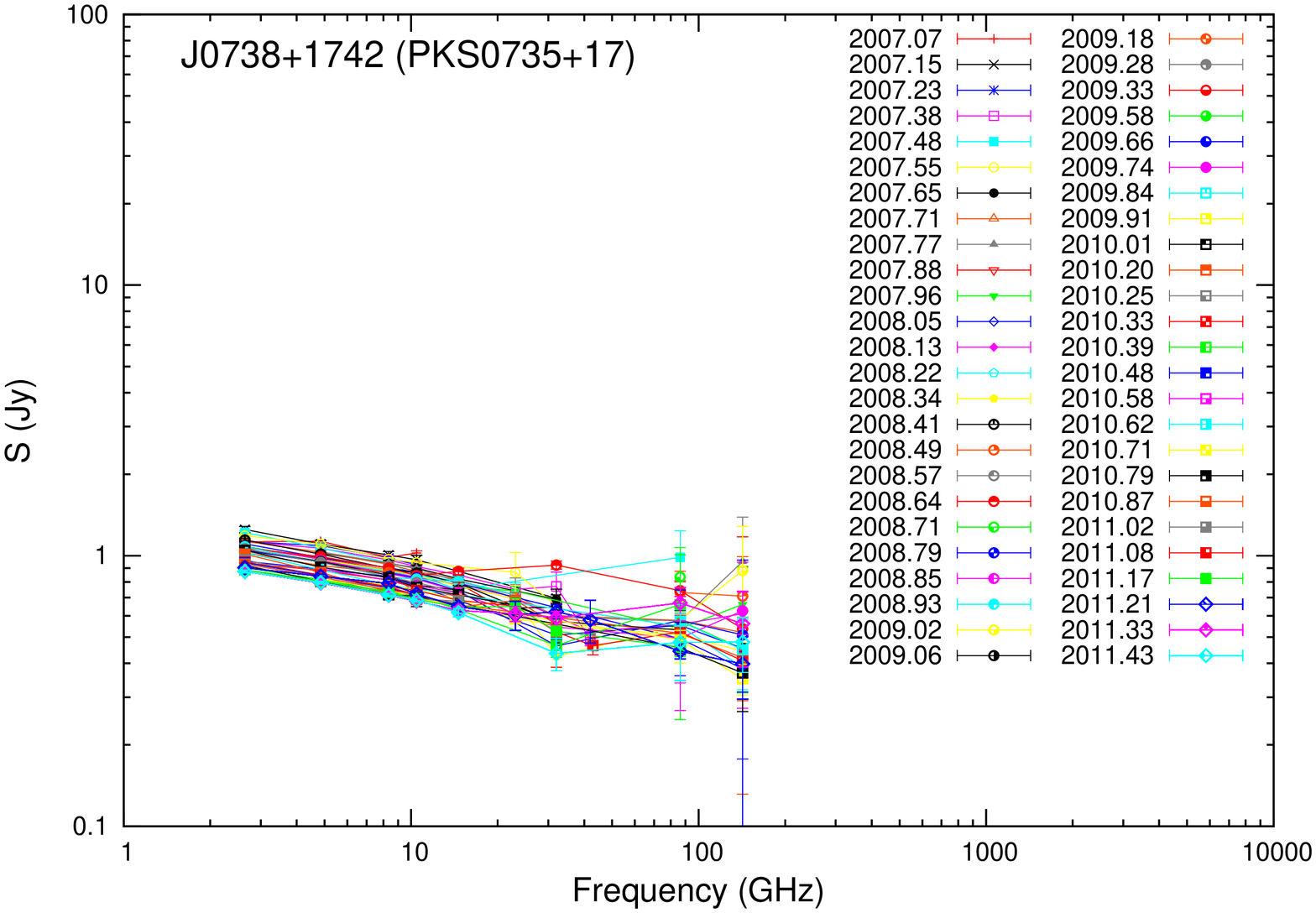}
\caption{\label{fig:t4b}Prototype source for variability type 4b.}
\end{minipage}\hspace{1pc}\vspace{1pc}%
\begin{minipage}{12pc}
\includegraphics[width=0.67\textwidth,angle=-90]{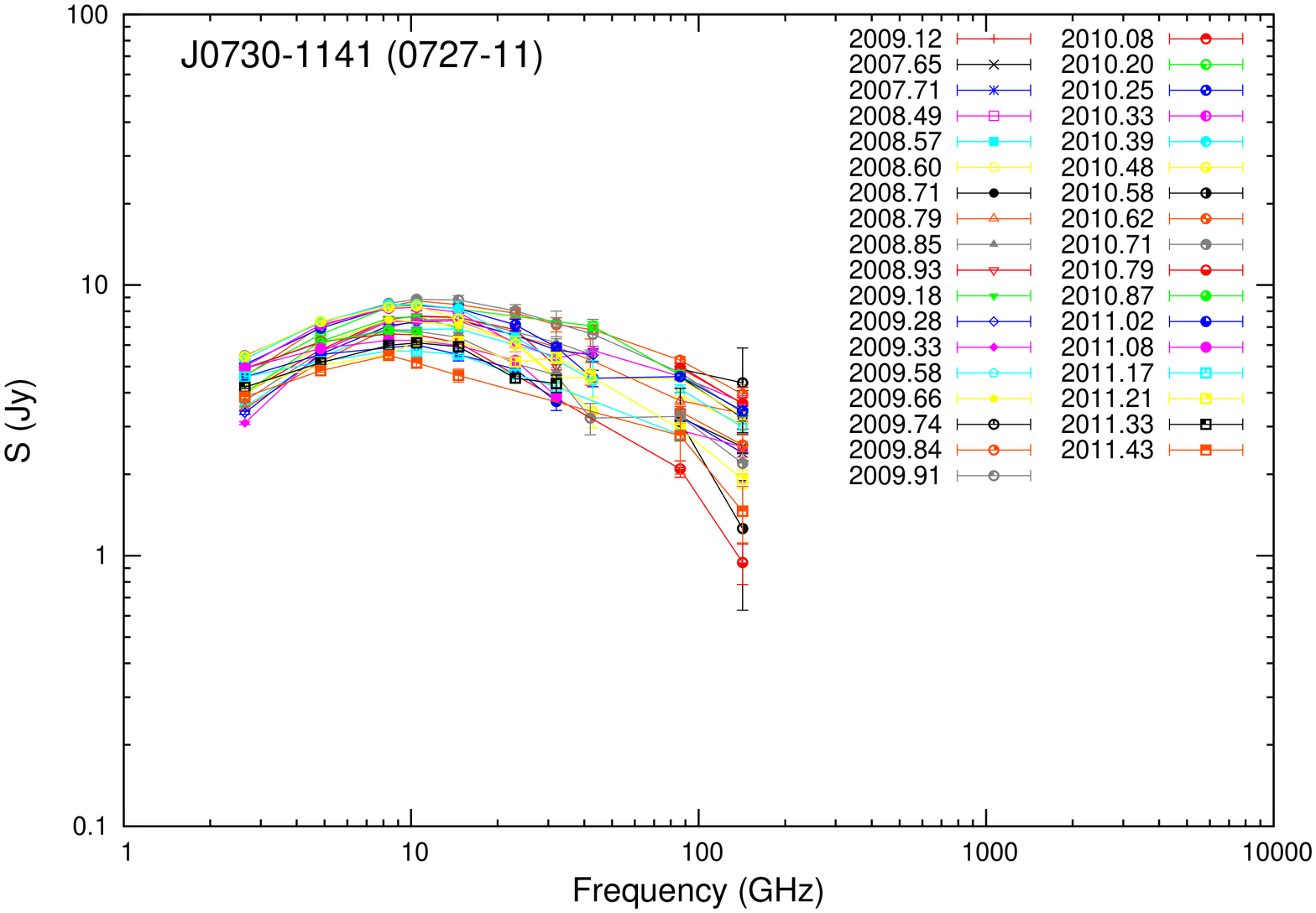}
\caption{\label{fig:t5}Prototype source for variability type 5.}
\end{minipage}\hspace{1pc}%
\begin{minipage}{12pc}
\includegraphics[width=0.67\textwidth,angle=-90]{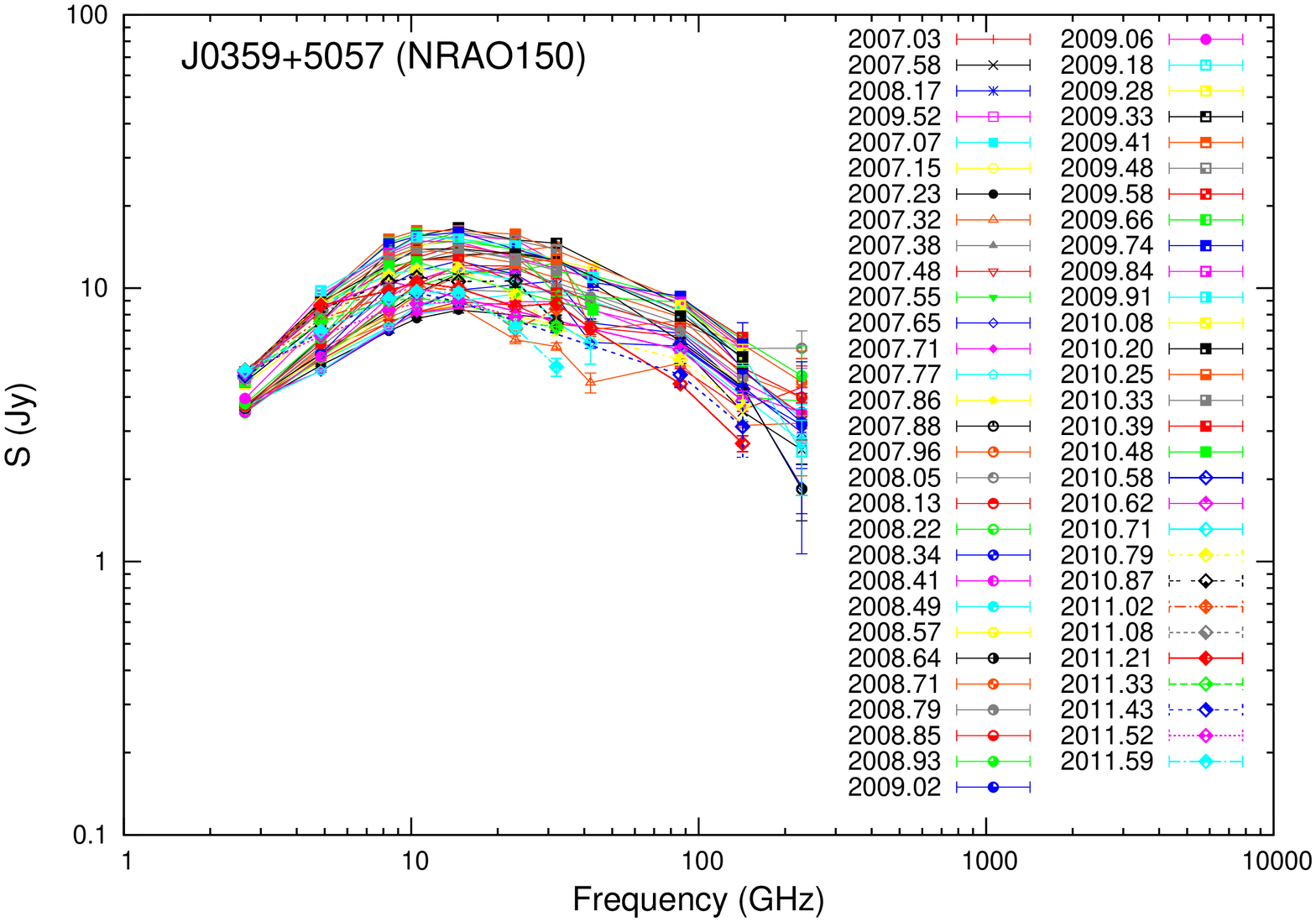}
\caption{\label{fig:t5b}Prototype source for variability type 5b.}
\end{minipage} 
\end{figure*}

One of the most comprehensive radio monitoring efforts, the {\em F-GAMMA} program,
\citep{fuhrmann2007AIPC,angelakis2008MmSAI..79.1042A,2010arXiv1006.5610A} was initiated in
2007 utilising the Effelsberg 100-m, the IRAM 30-m and the APEX 12-m telescopes for the
monthly monitoring of $\sim60$ {\sl Fermi} blazars at 12 frequencies between
2.6 and 345\,GHz. The millimetre observations are closely coordinated with the more
general flux monitoring conducted by IRAM at the 30-m telescope. The Effelsberg 100-m
telescope is equipped with circularly polarised feeds, while the IRAM 30-m with linearly
polarised ones. Details are given elsewhere (Fuhrmann et al. in prep., Angelakis et al. in
prep., Nestoras et al. in prep.).  Measurements at 4.85\,GHz, 10.45\,GHz, 32.0\,GHz,
86.24\,GHz and 142\,GHz are done differentially either by using multi-feed systems, or, at
IRAM 30-m, by wobbler switching. On average, the time needed for observing an entire
spectrum of any given source at Effelsberg alone is of the order of 35 minutes, while at
IRAM, roughly 2 minutes. The combined spectra (Effelsberg and IRAM) are observed
quasi-simultaneously within approximately one week. That is, neither the single-facility
spectra nor the combined ones are likely to be affected by source variability. In the
current study only data collected until June 2011, have been used.


\section{BROAD-BAND RADIO SPECTRA VARIABILITY}
The data product of the {\em F-GAMMA} program is monthly sampled broad-band radio spectra
covering a frequency range of 2 orders of magnitude. In the following we discuss studies
based only on Effelsberg and IRAM data i.e. 2.6 - 142\,GHz and present a toy-model for the
interpretation of the small number of phenomenological types of variability patterns.

\subsection{Phenomenological Classification}
\label{classes}
\citet{2011arXiv1111.6992A,2012arXiv1202.4242A} showed that the variability patterns that the source spectra
follow as a function of time, fall in a small number of phenomenological classes the
prototypes of which are shown in Figures~\ref{fig:t1}~-~\ref{fig:t5b}. A detailed
description of the characteristics of each type are given in
\cite{2011arXiv1111.6992A,2012arXiv1202.4242A}. What is readily obvious in this analysis and can also be seen
in Figures~\ref{fig:t1}~-~\ref{fig:t5b}, is that the patterns can be basically classified
in (a) those characterised by spectral evolution (Figures~\ref{fig:t1}~-~\ref{fig:t4b})
and (b) those that show a convex spectrum which varies self-similarly following a
close-to-achromatic evolutionary path (Figures~\ref{fig:t5} and \ref{fig:t5b}).

\subsection{Interpretation of the Observed Phenomenologies}
\begin{figure}[]
\centering
\includegraphics[width=0.38\textwidth]{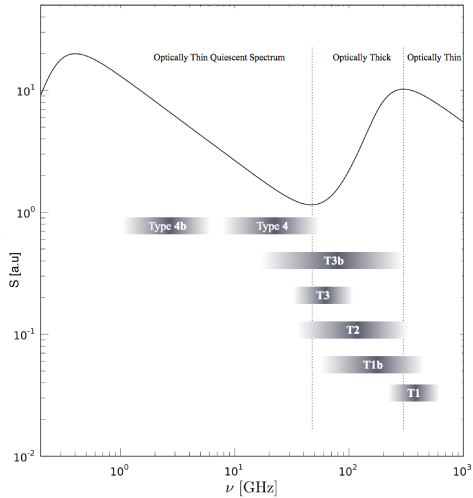}
\caption{The assumed two-component system. The different variability types can be
  reproduced with the appropriate modulation of the relative position and relative
  broadness of the band-pass denoted by the grey shaded areas.} \label{fig:principal}
\end{figure}
As it is argued in \cite{2011arXiv1111.6992A}, the spectral-evolution dominated types can
naturally be explained by a simple two-component system, as it is observed by other
programs as well \cite[e.g.][]{2002PASA...19...83K}. It is shown that a system composed of
a steep spectrum component attributed to a large scale jet populated with a varying
spectral component, can easily reproduce the observed patterns
(Figure~\ref{fig:principal}). The spectral component is assumed to be following the
\cite{Marscher1985ApJ} model. Calculations showed that with only a minor coverage of the
parameter space that determines the characteristics of those two elements (e.g. intrinsic
source luminosity, flare dominance, turnover frequency of quiescent spectrum), is enough
to reproduce the observed variability.  This discussion refers only to these types that
are characterised by spectral evolution. The ``achromatic'' behaviour must be attributed
to different mechanism as it will be discussed elsewhere.

\subsection{Variability Properties}
\begin{figure*}[]
\begin{minipage}{12pc}
\includegraphics[width=0.77\textwidth,angle=-90]{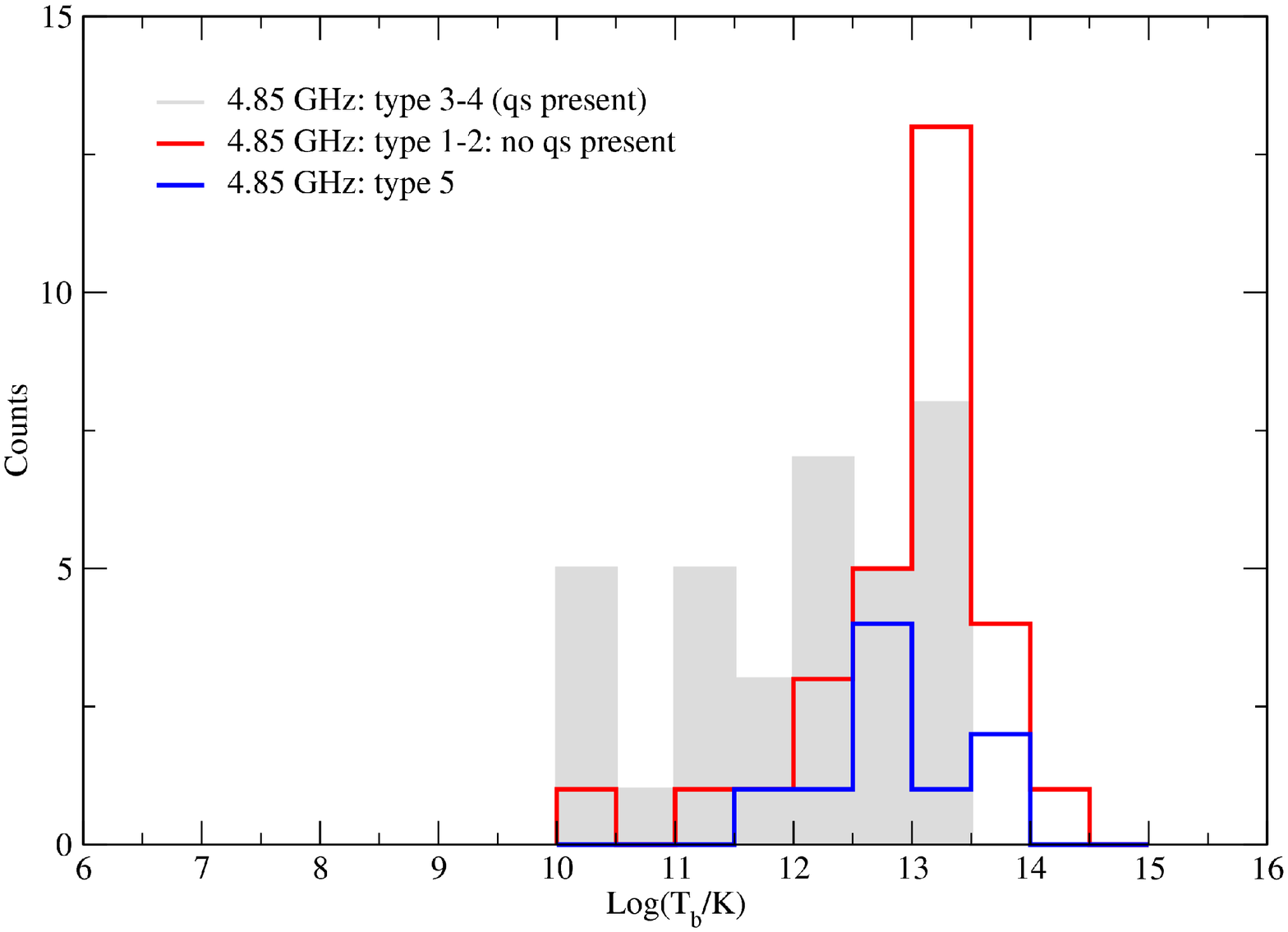}
\end{minipage}\hspace{1pc}\vspace{1pc}%
\begin{minipage}{12pc}
\includegraphics[width=0.77\textwidth,angle=-90]{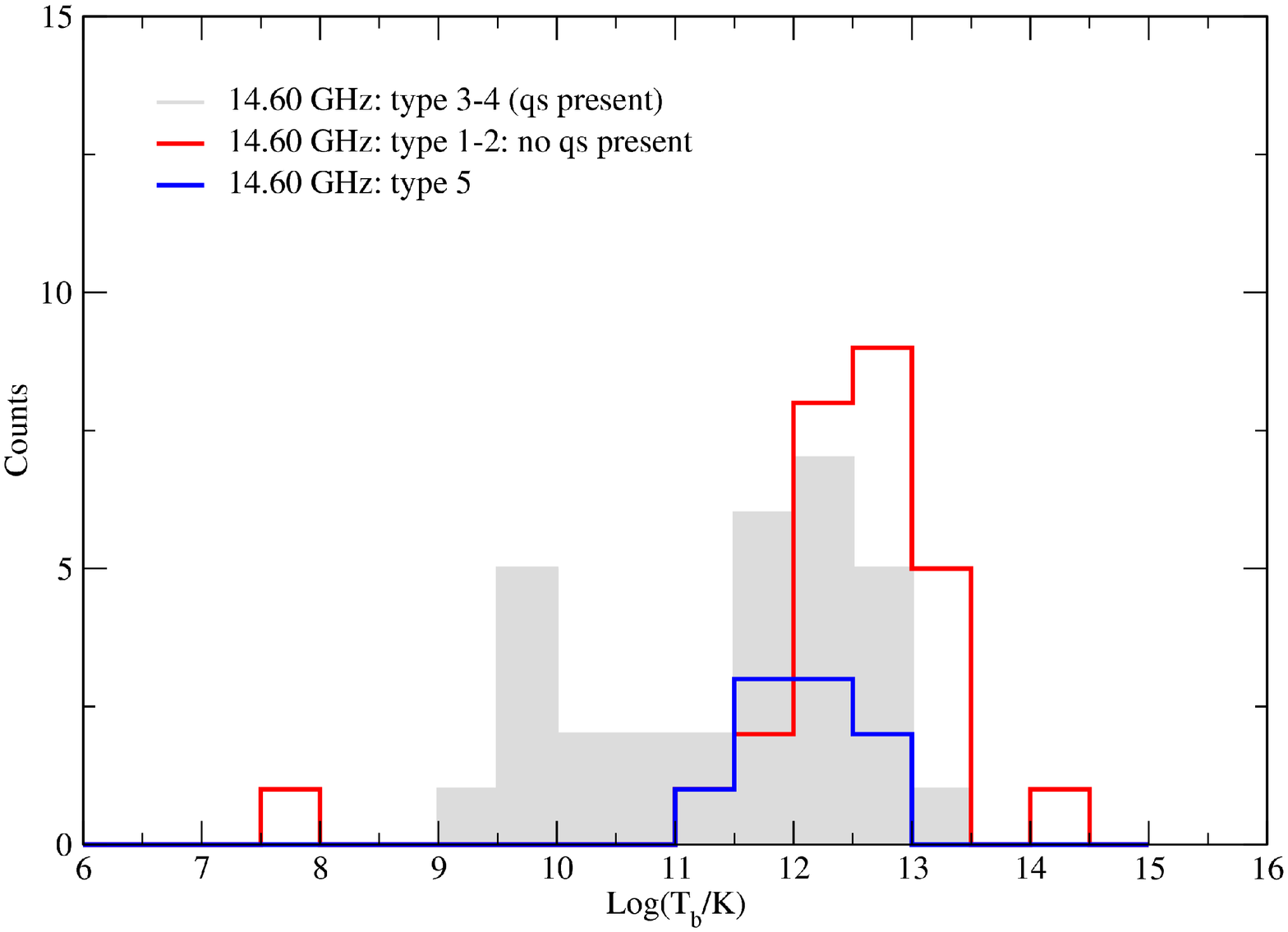}
\end{minipage}\hspace{1pc}%
\begin{minipage}{12pc}
\includegraphics[width=0.77\textwidth,angle=-90]{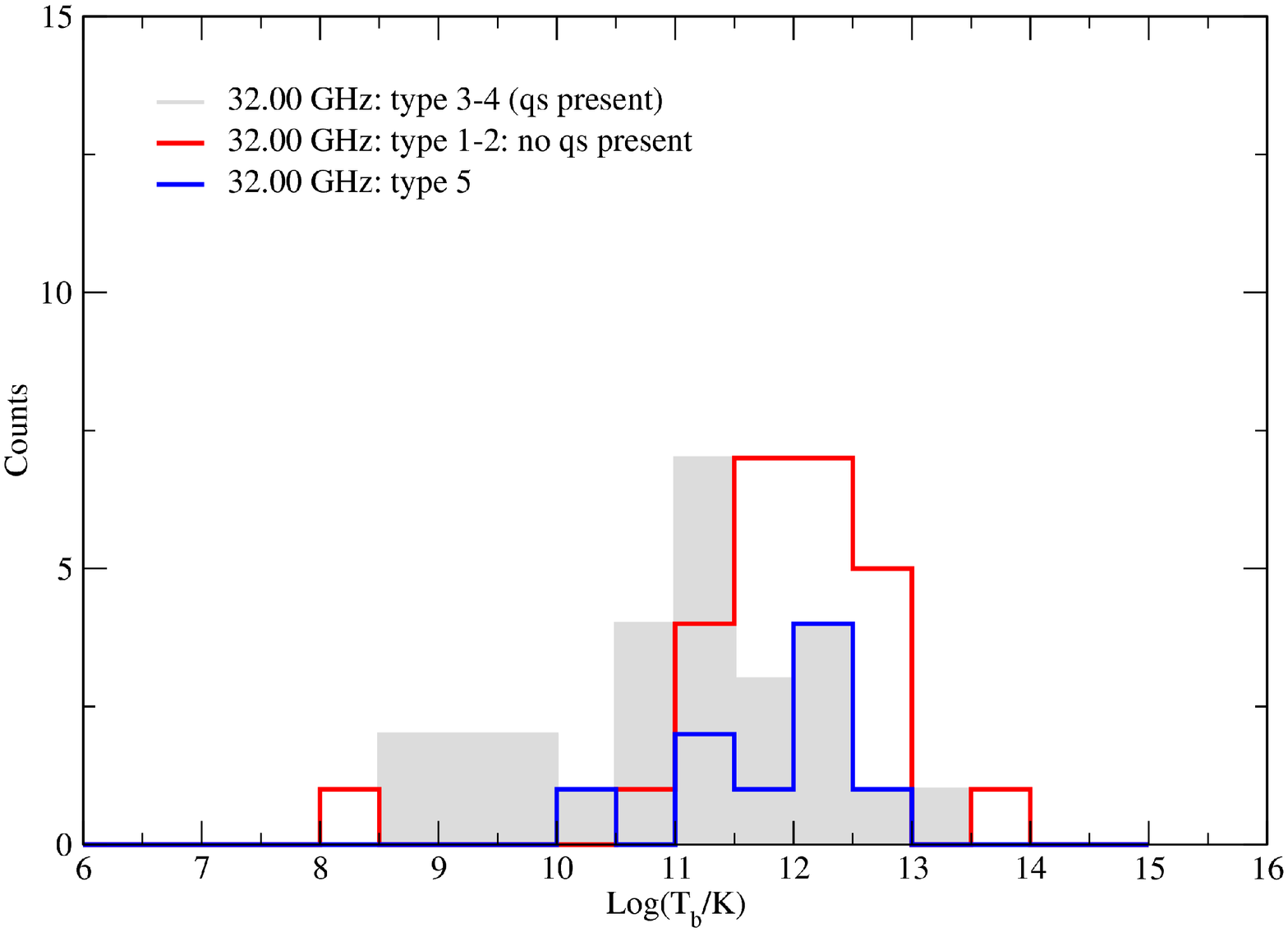}
\end{minipage}\hspace{1pc}%
\begin{minipage}{12pc}
\includegraphics[width=0.77\textwidth,angle=-90]{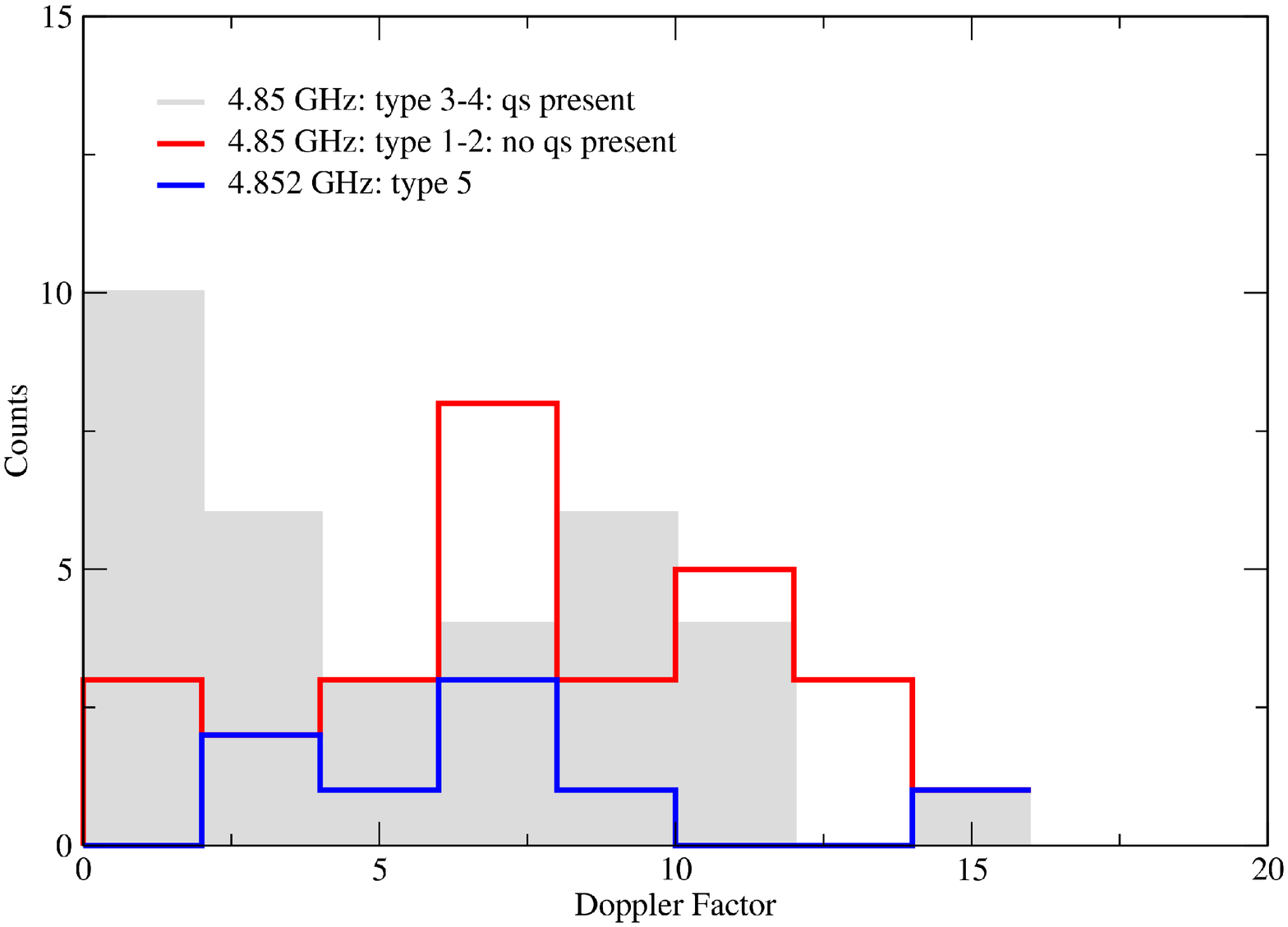}
\end{minipage}\hspace{1pc}\vspace{1pc}%
\begin{minipage}{12pc}
\includegraphics[width=0.77\textwidth,angle=-90]{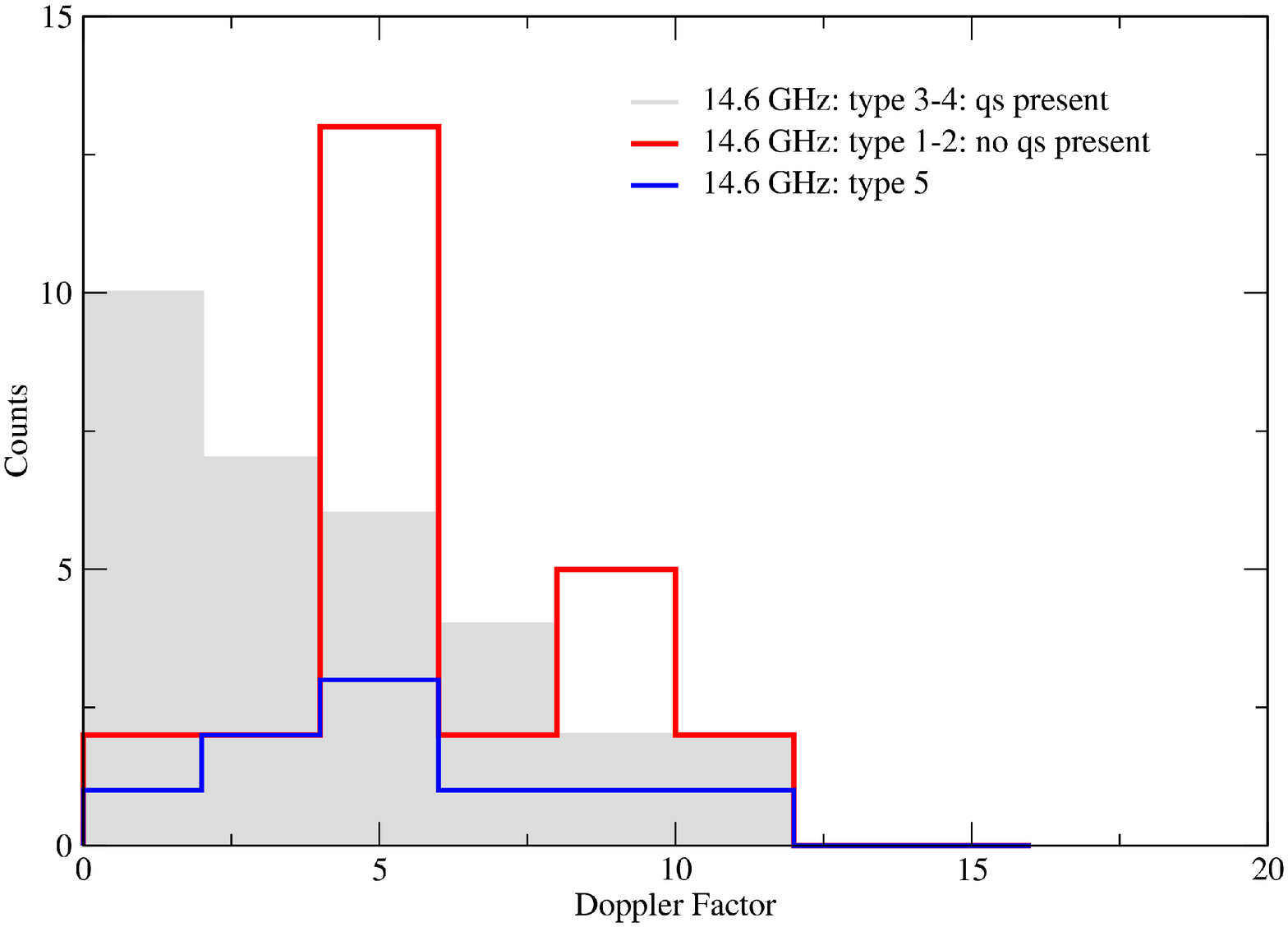}
\end{minipage}\hspace{1pc}%
\begin{minipage}{12pc}
\includegraphics[width=0.77\textwidth,angle=-90]{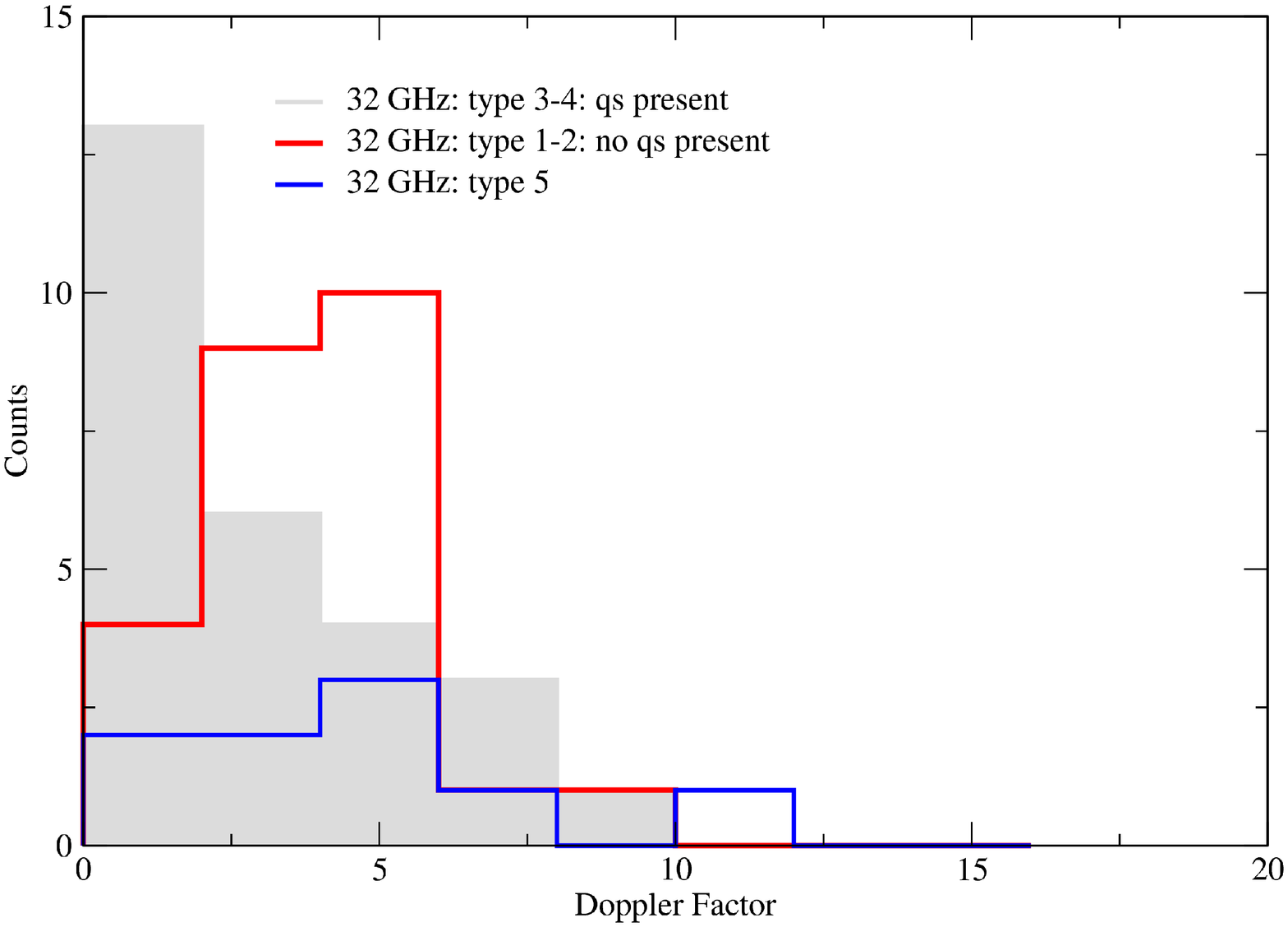}
\end{minipage}\hspace{1pc}%
\caption{\label{fig:Tb}Variability brightness temperatures and the inferred equipartition
  Doppler factor distributions. {\sl Red} denotes sources of variability type 1-2 that
  show no clue of quiescent spectrum. {\sl Grey} denotes sources where there is an
  evident quiescent spectrum and {\sl blue} are sources of type 5. {\sl Left:} 4.85\,GHz,
  {\sl middle:} 14.6\,GHz, {\sl right:} 32\,GHz.}
\end{figure*}
The different classes discussed earlier have some very fundamental differences. The most
obvious one being the increase in the prominence of a quiescent large scale jet as compared
to the flaring event with increasing type. That should immediately imprint differences in
the distribution of the variability parameters for types that have different quiescent
spectrum prominence.

In order to investigate that, we have performed standard variability analysis. For each
light curve characteristic variability time scales have been estimated using a structure
function analysis described by \cite{Simonetti1985ApJ}. These time scales are subsequently
used to compute the involved emitting region sizes (assuming the light-travel-time
argument) and hence the variability brightness temperatures, as:
\begin{equation}\label{eq:Tb}
T_\mathrm{b}\mathrm{[K]}=4.5\cdot10^{10}\Delta S \left(\frac{\lambda \cdot D_\mathrm{L}}{\Delta\tau \cdot(1+z)^2}\right)^2
\end{equation}
\begin{tabular}{rrl}
  where &$\Delta S$:& the flux density variation in Jy\\
  &$\lambda$:& the observing wavelength in cm\\
  &$D_\mathrm{L}$:& the luminosity distance in Mpc\\
  &$\Delta\tau$:& the characteristic time scale in days\\
\end{tabular}\\\\

Assuming a typical value of $5\cdot10^{10}$\,K for the upper limit in the brightness
temperature on the basis of equipartition arguments \citep{Readhead1994ApJ,1999ApJ...511..112L}, we can
compute also the variability equipartition Doppler factor, as:
$D_\mathrm{var}=(1+z)\cdot\left(T_\mathrm{b}/
  5\cdot10^{10}\mathrm{\,K}\right)^\frac{1}{3-\alpha}$, where $\alpha$ is the spectral
index ($S\propto\nu^{\alpha}$). A typical value of $\alpha=-0.7$ has been used.

In Figure~\ref{fig:Tb} we show the variability brightness temperature (Eq.~\ref{eq:Tb})
distribution at three characteristic frequencies, 4.85, 14.6 and 32\,GHz, for three groups
of sources separately. One group is made of sources of type 3-4 which show the clear
presence of a quiescent spectrum (qs); sources of type 1-2 comprise the second group and
show no evidence for the presence of such a spectrum.  The last group is made of sources
that belong to type 5. In the lower row of Figure~\ref{fig:Tb} we show the same
distributions for the variability equipartition Doppler factors, D.

From these plots it can be seen that at all bands there is a separation between the
distribution of the first group when compared with the other two ones. This could be
explained as follows; for sources of type 1-2 or 5 the dominance of flares relative to the
quiescent spectrum is significant. Given the fact that the brightness temperature and the
corresponding Doppler factor are inferred from the variability properties, it is naturally
expected that sources where the quiescent spectrum is absent, show more pronounced
variability characteristics. At the highest frequencies this separation must disappear
due to the absence of quiescent spectrum.

\section{JET EMISSION FROM NLSy1s}
Narrow Line Seyfert 1 galaxies are an AGN sub-class that have been argued to show
comparably low masses and high accretion rates
\citep[e.g.][]{2011nlsg.confE...3B,2011nlsg.confE...4G,2011nlsg.confE..35M,2011arXiv1109.4181P}. Only
a mere 7\,\% of them appears to be radio loud \citep{2006AJ....132..531K}.

The early {\sl Fermi} discovery of $\gamma$-ray emission from a small number of NLSy1s
\citep{2009ApJ...707L.142A,2009ApJ...699..976A}) came as surprise; until then the only
$\gamma$-ray bright classes were thought to be blazars (i.e. FSRQs and BL Lac objects) and
radio galaxies (for a review, see \cite{2011nlsg.confE..24F}).  The discovery of
$\gamma$-ray emission from NLSy1s not only introduced a new class of $\gamma$-emitting AGNs but
also revolutionised the well spread belief that jets are associated (chiefly) with large
elliptical galaxies. The multi-wavelength campaign between March and July 2009 following
the discovery of $\gamma$-ray emission from PMN\,0948$+$0022 showed that the source was
exhibiting a spectral behaviour typical of a relativistic jet
\citep{2009ApJ...707..727A,2011AnA...528L..11G}.
\begin{table}[]
\begin{center}
\caption{The NLSy1 monitored by the {\em F-GAMMA} program.}
\begin{tabular}{lcccc}
  \hline \textbf{Source} & {\bf $z$} & \textbf{$N^\dagger$} &\textbf{$f^\ddagger$} &{\bf $\left<S_{4.85}\right>$}\\
  &                &       &(months) &(Jy)\\
  \hline 
  1H\,0323$+$342   &0.061  &18 &1.0 &0.40 \\
  PMN\,J0948$+$0022&0.585 & 42 &0.9 &0.23 \\
  PKS\,1502$+$036  &0.409 &18 &0.9 &0.52 \\
  \hline
  \multicolumn{5}{l}{$^\dagger N$: number of observed epochs until February 2012}\\
  \multicolumn{5}{l}{$^\ddagger f$: mean sampling} 
\end{tabular}
\label{tbl:nlsy1s}
\end{center}
\end{table}

Responding to the $\sl Fermi$ detection the {\em F-GAMMA} team initiated a dedicated
program, to understand their cm to mm behaviour \citep{2011nlsg.confE..26F}. The three
sources that are monitored are listed in Table~\ref{tbl:nlsy1s}. In the following we
discuss first results of this monitoring.

\subsection{Radio Variability}
\begin{figure*}[]
\begin{minipage}{12pc}
\includegraphics[width=0.77\textwidth,angle=-90]{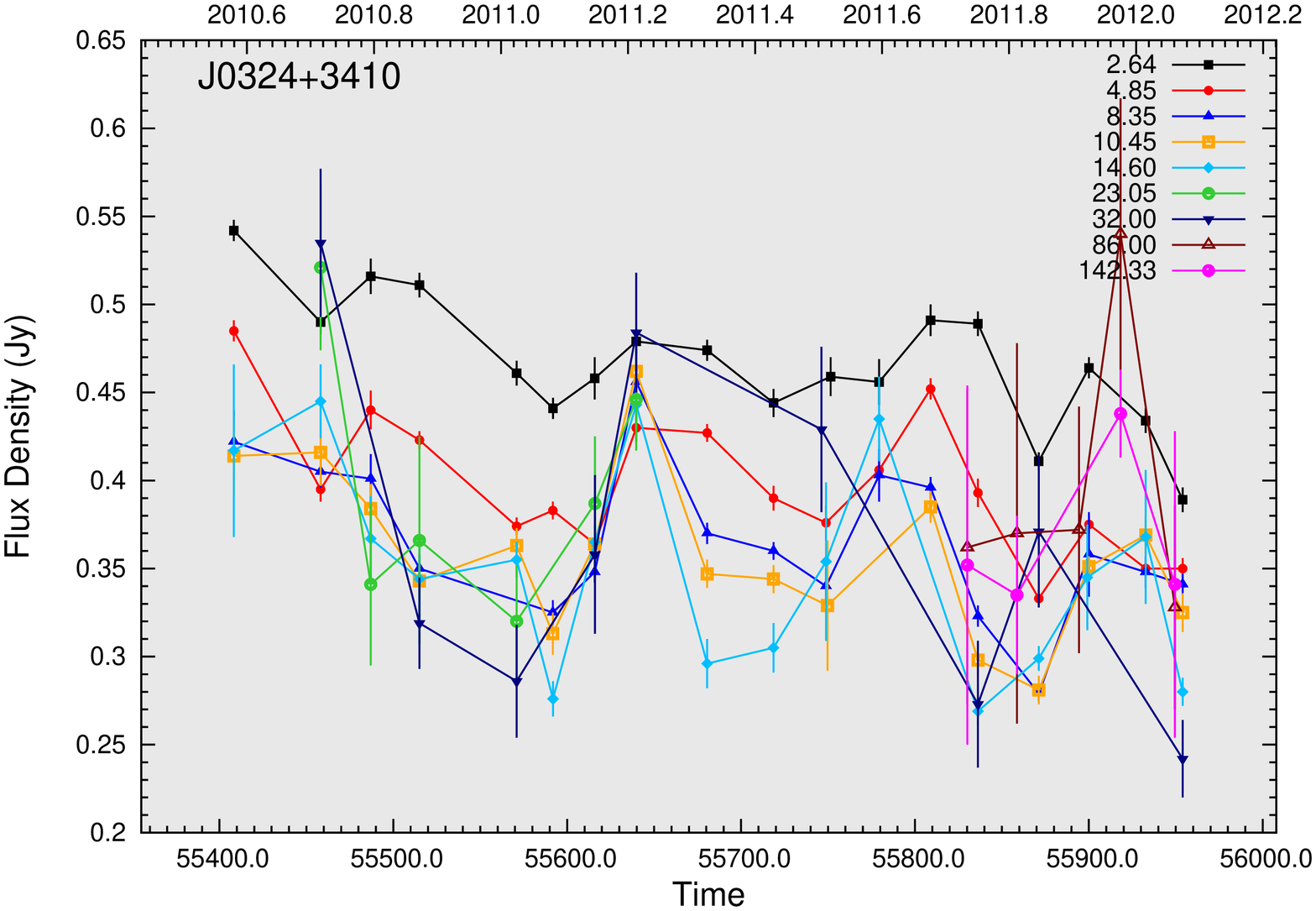}
\end{minipage}\hspace{1pc}\vspace{1pc}%
\begin{minipage}{12pc}
\includegraphics[width=0.77\textwidth,angle=-90]{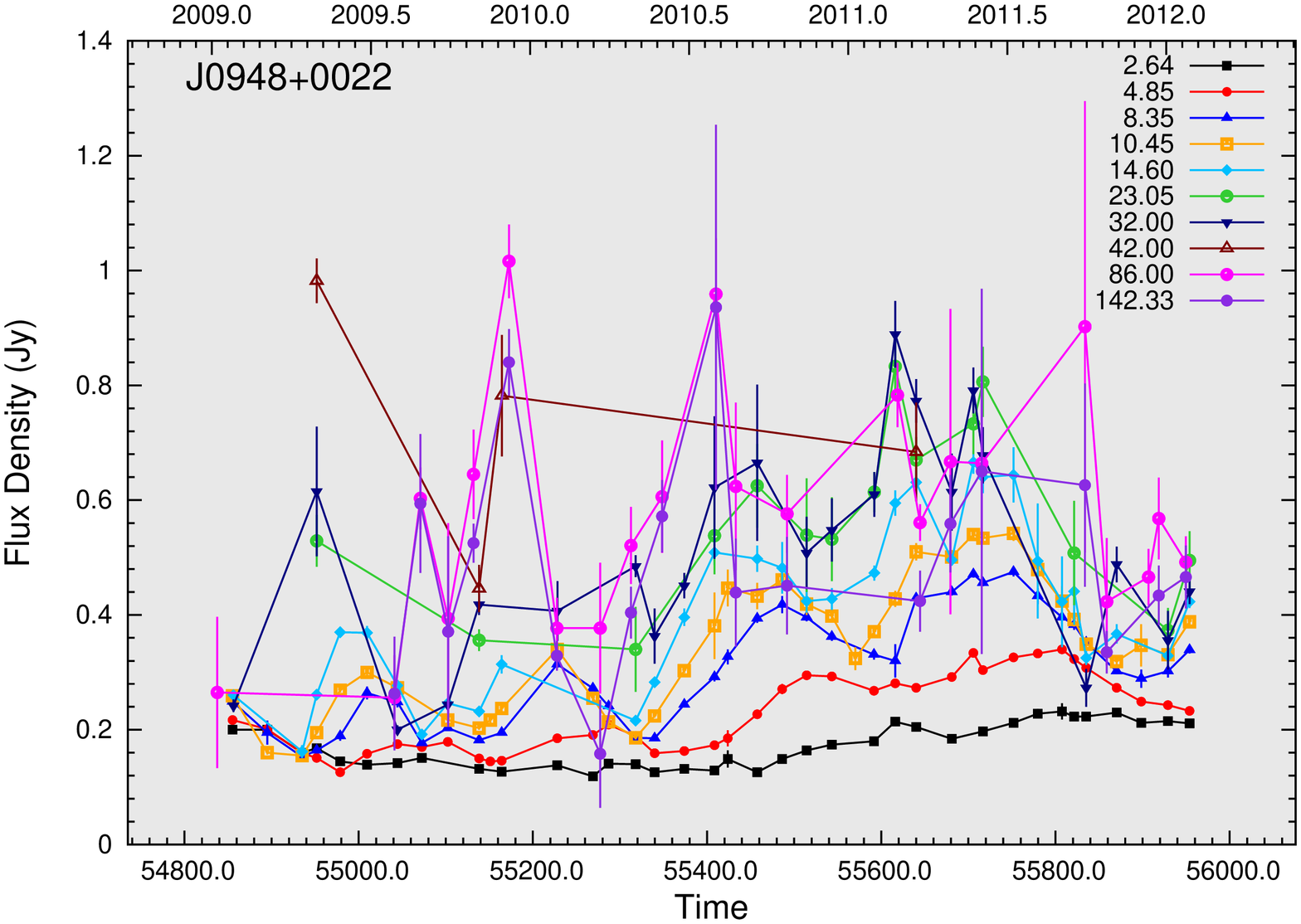}
\end{minipage}\hspace{1pc}%
\begin{minipage}{12pc}
\includegraphics[width=0.77\textwidth,angle=-90]{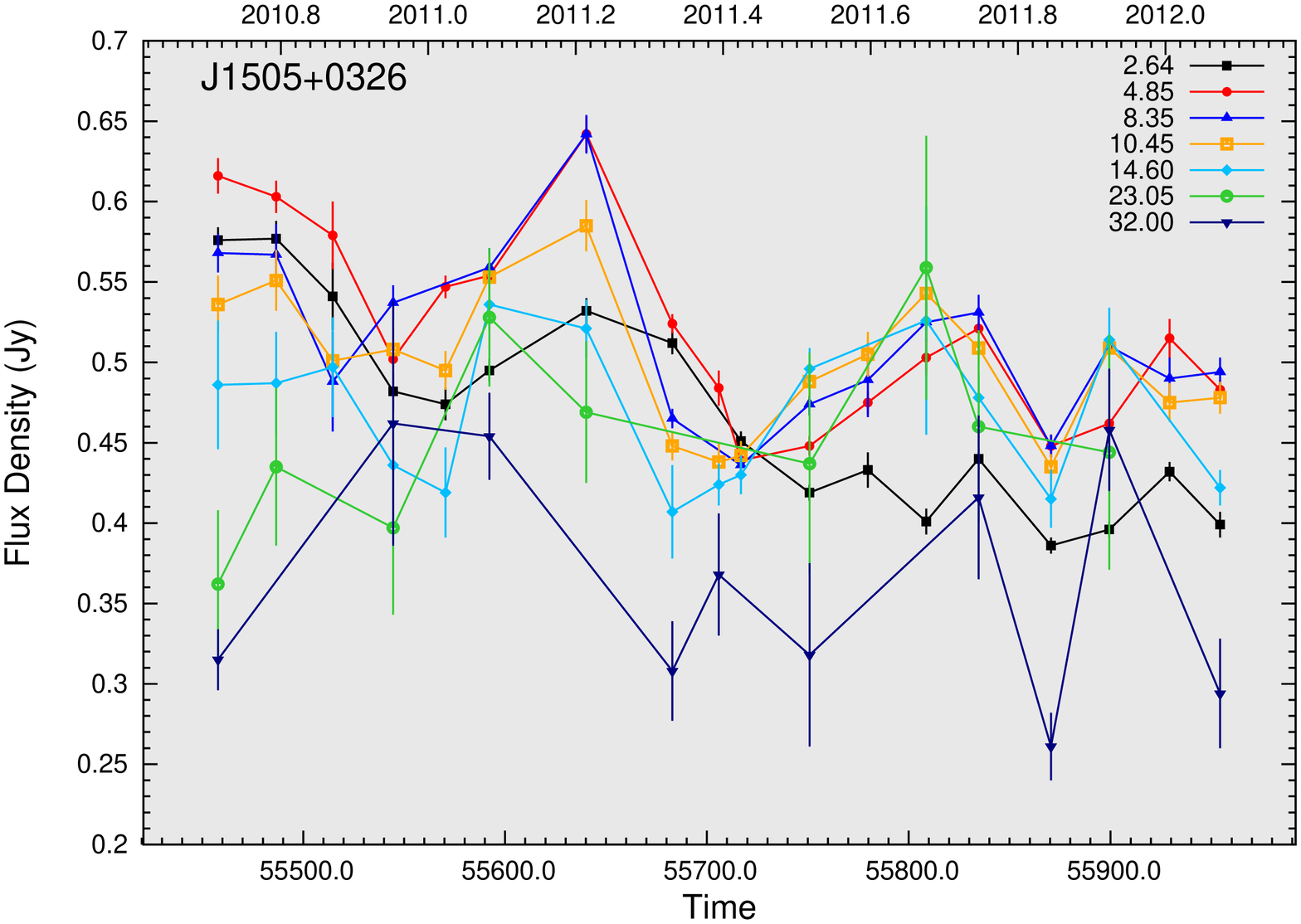}
\end{minipage}\hspace{1pc}%
\begin{minipage}{12pc}
\includegraphics[width=0.77\textwidth,angle=-90]{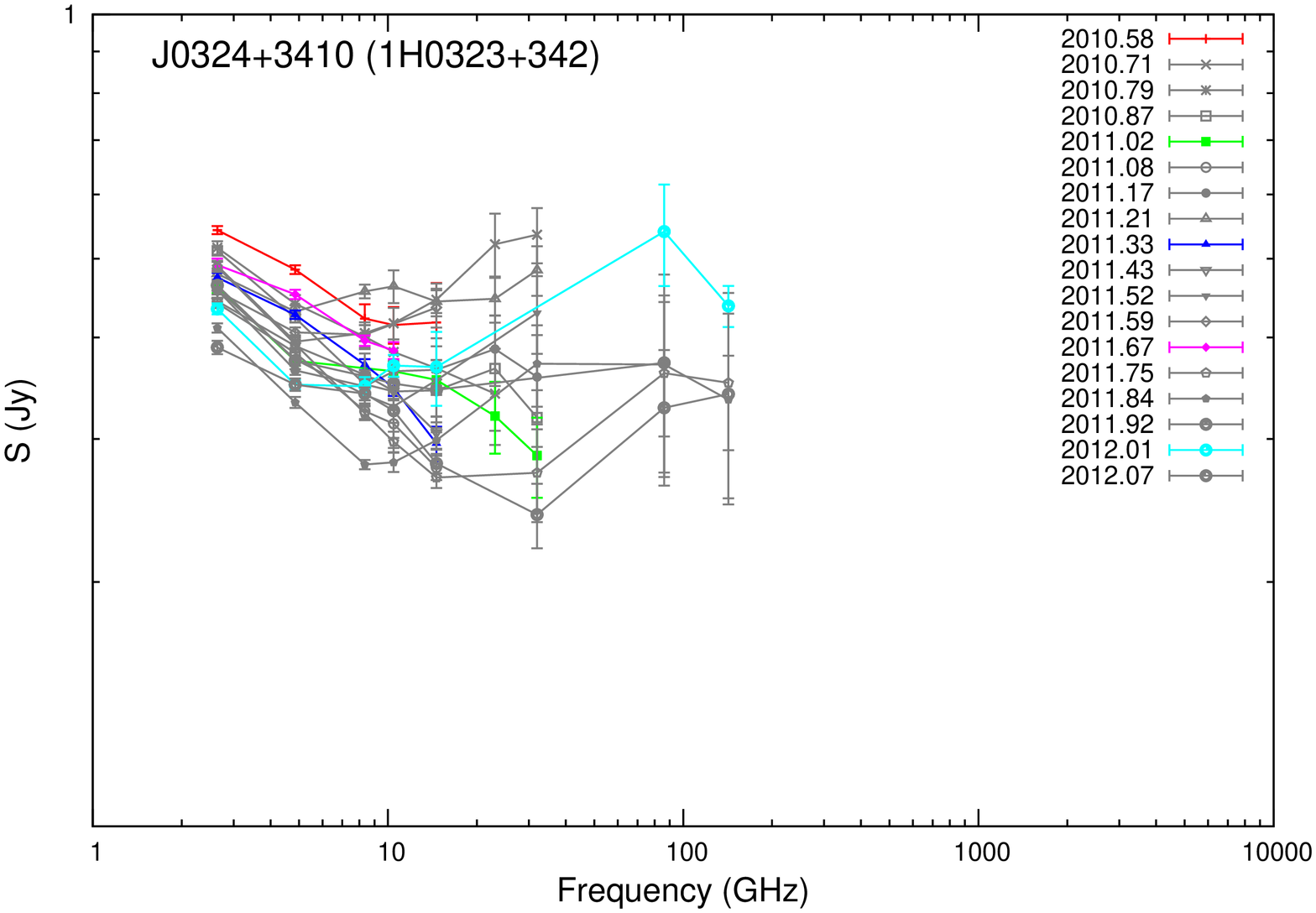}
\end{minipage}\hspace{1pc}\vspace{1pc}%
\begin{minipage}{12pc}
\includegraphics[width=0.77\textwidth,angle=-90]{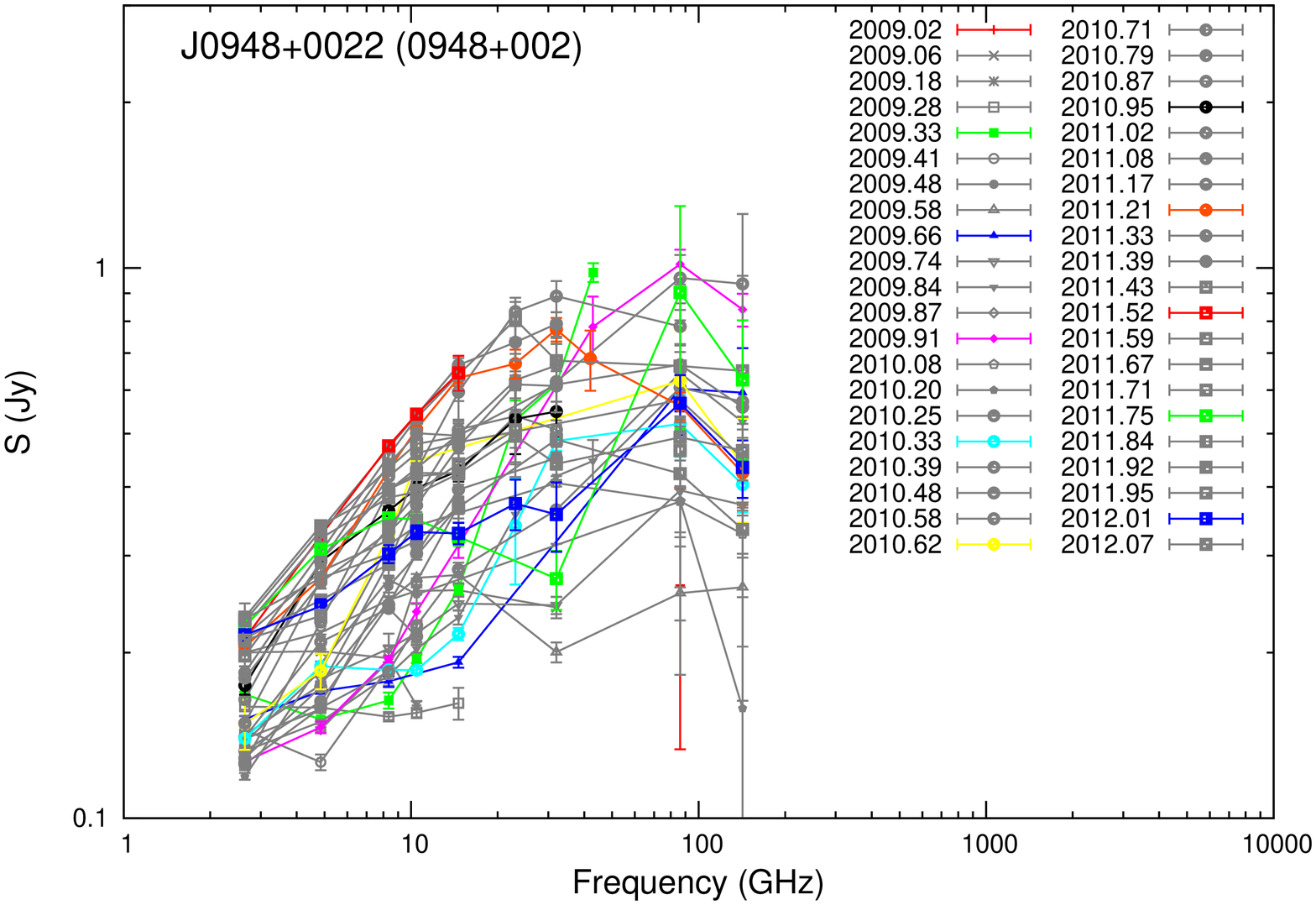}
\end{minipage}\hspace{1pc}%
\begin{minipage}{12pc}
\includegraphics[width=0.77\textwidth,angle=-90]{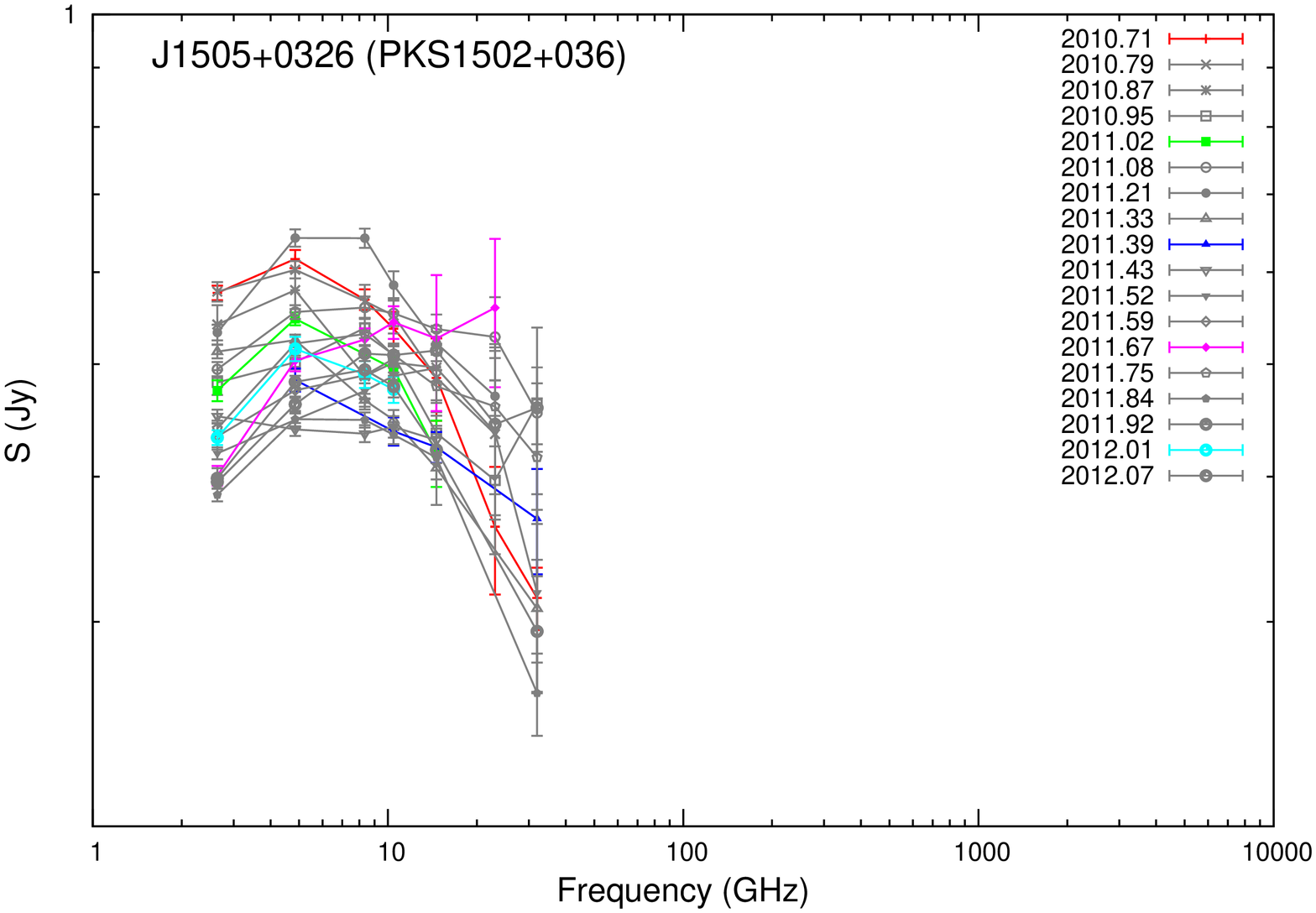}
\end{minipage}\hspace{1pc}%
\caption{\label{fig:LC-nlsy}Monthly sampled spectra and light curves for the three
  monitored NLSy1s. {\sl Upper row:} the light curves at all {\em F-GAMMA} frequencies between
  2.64 and 142\,GHz. {\sl Lower row:} the broad-band {\em F-GAMMA} spectra.}
\end{figure*}

In Figure~\ref{fig:LC-nlsy} we show the variability light curves as well as the broad-band
spectra observed within the context of the {\em F-GAMMA} program. For 1H\,0323$+$342 and
PMN\,J0948$+$0022 light curves between 2.6 and 142\,GHz have been constructed while the
steepening of the spectrum above 32\,GHz in the case of PKS\,1502$+$036, does not allow 
IRAM observations at short-mm bands.   
\\\\
{\bf 1H\,0323$+$342:} As it can be seen in Figure~\ref{fig:LC-nlsy},
its radio spectrum displays a quiescent part reaching up to roughly
10\,GHz beyond which a high frequency component (hereafter HFC)
appears. Its mean spectral index is of the order of $-0.23$
(calculated between 2.6 and 10\,GHz), not very different from a
typical value of $-0.5$ ($S\propto \nu^{\alpha}$). The HFC shows
intense spectral evolution which gradually shifts its peak
progressively towards the steep spectrum component. The pace of the
evolution is remarkably fast causing significant displacements within
the month that typically separates two observations. The variability
pattern according to the classification discussed in
Sect.~\ref{classes}, is of type 4. Its light curves although only
shortly sampled, shows a collection of events more prominent at higher
frequencies and with cross-band lags indicative of the spectral
evolution superimposed on a long term decreasing trend. The modulation
index $m$ ($m[\%]=100\times\frac{\sigma}{<S>}$) at 4.85\,GHz is,
$m_{4.85}\approx 10\,\%$, while at 14.6 and 32\,GHz it is, $17\,\%$
and $27\,\%$, respectively. The {\em Structure Function} analysis
applied on the source light curves revealed characteristic times
scales of the order of 60 days which implies a variability brightness
temperature of $10^{12}$\,K at 4.85\,GHz and $2\times10^{11}$\,K at
14.6\,GHz. The corresponding equipartition Doppler factors are 2.4 and
1.5 respectively, placing the source in the lower part of the Doppler
factor distribution shown in Figure~\ref{fig:LC-nlsy}. Interestingly,
the Doppler factor calculated from fitting the Spectral Energy
Distribution (SED) \citep{2009ApJ...707L.142A}
is around 17.\\\\
{\bf PMN\,J0948$+$0022:} It is evident from the light curves and the
spectra shown in Figure~\ref{fig:LC-nlsy} that the available dataset
for PMN\,J0948$+$0022 is much richer than that for the other two
NLSy1s. The spectrum appears mostly inverted representative of
variability type 1. The spectral index below 10\,GHz ranges between
marginally steep or flat ($\approx -0.1$) to highly inverted reaching
values of $+1.0$. Its evolution is exceptionally dynamic with
significant evolution happening even within one month. It appears that
for this source a sampling of two weeks would be necessary. Its light
curve shows at least 4 prominent events which emerge with time lags at
different bands. At the lowest frequencies the events are barely
seen. Yet, the modulation index shows a monotonic increase with
frequency. At 2.6, 4.85, 14.6, 32 and 142\,GHz the modulation index is
22, 29, 35, 37 and 38\,\%, respectively. The standard variability
analysis reveals brightness temperatures of $8\times10^{12}$\,K at
4.85\,GHz, $2\times10^{12}$\,K at 14.6\,GHz and $1.5\times10^{11}$\,K
at 32\,GHz which would imply rather typical Doppler factors of 6, 4
and 2 respectively, as it can be seen from the Doppler factor
distribution plots in Figure~\ref{fig:Tb}. The SED modelling gives Doppler factors that vary
between 10 and 20 with the latter being observed during the outburst
of July 2010
\cite{2011MNRAS.413.1671F}. \\\\
{\bf PKS\,1502$+$036:} This source shows a variability behaviour
similar to type 1b although more epochs are needed for a definite
classification. Its spectrum is highly variable displaying periods of
convex shape. Its low-band part ($\nu\le 8$\,GHz) varies between flat
and highly inverted, while the higher frequencies
($10\mathrm{\,GHz}\le \nu$) can show spectral index as steep as
$-0.5$. The light curve shows at least 2 events better seen at higher
frequencies. The high frequency cut-off of the spectrum prohibits IRAM
monitoring. The typical time scales identified here are of the order
of 60 - 80 days. At 4.85\,GHz the brightness temperature is
$2\times10^{13}$\,K and at 14.6\,GHz it is $3\times10^{12}$\,K
implying Doppler factors of 7 and 4, respectively while the SED
fitting gives a value of 18.

\subsection{Conclusions}
From the above discussion it becomes obvious that the three monitored
NLSy1s show a very typical blazar-like behaviour. That is, highly
variable spectra caused by the presence of prominent evolving high
frequency spectral components. The variability happens at
interestingly fast pace with the mean number of events per unit time
being clearly larger than that of the rest of the {\em F-GAMMA}
targets. As an example, PMN\,J0948$+$0022 shows, on average, 1.4
flaring events per year as compared to less than $\sim 1$ event per
year for other typical {\em F-GAMMA} blazars.  The variability
brightness temperature is relatively high with respect to the
distribution of the whole sample in the cases of PMN\,J0948$+$0022 and
PKS\,1502$+$036. The opposite is the case for 1H\,0323$+$342. The
derived Doppler factors are lower limits and are systematically lower
than the values obtained from the SED modelling. In any case, it seems
premature to draw any final conclusions for the behaviour of this
limited sample of NLSy1s. Longer time baselines will allow more
accurate estimates of the brightness temperatures and all the
variability parameters.

\bigskip 
\begin{acknowledgments}
  Based on observations with the 100\,m telescope of the MPIfR (Max-Planck-Institut f\"ur
  Radioastronomie) and the IRAM 30m Telescope. IRAM is supported by INSU/CNRS (France),
  MPG (Germany) and IGN (Spain).  I. Nestoras and R. Schmidt are members of the
  International Max Planck Research School (IMPRS) for Astronomy and Astrophysics at the
  Universities of Bonn and Cologne. E. Angelakis wholeheartedly thanks Dr. A. Kraus for
  the constant support and the constructive discussions and Dr. T. Savolainen for all the
  usefull comments. Finally, E. Angelakis thanks the LOC and the SOC of the {\em ``Fermi and
    Jansky: Our Evolving Understanding of AGN''} conference, for organizing such an
  interesting meeting and the anonymous referee for the thorough
  reading of the paper and  the constructive comments.
\end{acknowledgments}

\bigskip 






\end{document}